# Spatial homogeneity and doping dependence of quasiparticle tunneling spectra in cuprate superconductors

N.-C. Yeh[a], C.-T. Chen[a], G. Hammerl[b], J. Mannhart[b], S. Tajima[c], K. Yoshida[c], A. Schmehl[b], C. W. Schneider[b], and R. R. Schulz[b]

[a] Department of Physics, California Institute of Technology, Pasadena, CA 91125, USA
[b] Center for Electronic Correlations and Magnetism, Institute of Physics, Augsburg University, Augsburg, Germany
[c] Superconductivity Research Laboratory, International Superconductivity Technology Center, Tokyo 135, Japan

Scanning tunneling spectroscopy (STS) studies reveal long-range (~$100$ $nm$) spatial homogeneity in optimally and underdoped superconducting $YBa_2Cu_3O_{7-\delta}$ (*YBCO*) single crystals and thin films, and *macroscopic* spatial modulations in overdoped $(Y_{0.7}Ca_{0.3})Ba_2Cu_3O_{7-\delta}$ (*Ca-YBCO*) epitaxial films. In contrast, STS on an optimally doped $YBa_2(Cu_{0.9934}Zn_{0.0026}Mg_{0.004})_3O_{6.9}$ single crystal exhibits strong spatial modulations and suppression of superconductivity over a *microscopic* scale near the *Zn* or *Mg* impurity sites, and the global pairing potential is also reduced relative to that of optimally doped *YBCO*, suggesting strong pair-breaking effects of the non-magnetic impurities. The spectral characteristics are consistent with $d_{x^2-y^2}$ pairing symmetry for the optimally and underdoped *YBCO*, and with $(d_{x^2-y^2}+s)$ for the overdoped *Ca-YBCO*. The doping-dependent pairing symmetry suggests interesting changes in the superconducting ground state, and is consistent with the presence of nodal quasiparticles for all doping levels. The maximum energy gap $\Delta_d$ is non-monotonic with the doping level, while the ($2\Delta_d/k_BT_c$) ratio increases with decreasing doping. The similarities and contrasts between the spectra of *YBCO* and of $Bi_2Sr_2CaCu_2O_{8+x}$ (*Bi-2212*) are discussed.

## 1. Unconventional pairing symmetry and strong phase fluctuations in cuprate superconductors

Cuprate superconductors are doped Mott insulators with physical properties differing significantly from those of conventional superconductors [1,2]. Some of the hallmarks of these cuprates include the marginal Fermi-liquid (MFL) behavior [3] and pseudogap phenomena [4] in the normal state of underdoped and optimally doped samples, the predominantly $d_{x^2-y^2}$ pairing symmetry in the superconducting state [5,6], and the small phase stiffness and large fluctuation effects [7,8]. Different mechanisms have been proposed to account for the novel physical properties of cuprates, such as the resonance valence bond (RVB) theory [9,10] and stripe-phase scenario [11,12] for spin-charge separation in the normal state, and the circular current phase and quantum criticality scenario [3] for the MFL behavior. Recently, it has been proposed that the interaction of nodal quasiparticles with thermally induced vortex loops [7] may give rise to the MFL behavior in the normal state and below the pseudogap temperature ($T^*$) [13]. Generally speaking, an important consequence of the *d*-wave pairing symmetry is the presence of nodal quasiparticles at low temperature, and that of strong phase fluctuations is the separation of the pair formation temperature ($T_{MF} \sim T^*$) from the superconducting transition temperature ($T_c$) [7], with $T_c << T^*$. Given the fact that the low-energy excitations are important manifestations of the pairing state, we expect nodal quasiparticles to play a major role in determining the physical properties of the cuprates. On the other hand, if the Fermi surface were fully gapped due to broken time-reversal ($\mathcal{T}$) symmetry in the pairing state, as suggested by certain theories [14,15], the low-energy excitation spectra at $T << T^*$ would have been modified significantly. It is therefore important to establish the purity and possible doping dependence of the pairing symmetry in the cuprate superconductors. In particular, whether a small broken $\mathcal{T}$-symmetry component may exist in the form of ($d_{x^2-y^2}+id_{xy}$) or ($d_{x^2-y^2}+is$) pairing should be carefully examined. In this work, we report studies of the quasiparticle tunneling spectra on the $YBa_2Cu_3O_{7-\delta}$ (*YBCO*) system over a wide range of doping levels. Several key results are noteworthy. First, we observed long-range (~ $100$ $nm$) spatial homogeneity in the pairing potential of both underdoped and optimally doped *YBCO*, and the pairing symmetry is consistent with $d_{x^2-y^2}$ within our experimental resolution. Second, *macroscopic* spatial modulation of the pairing potential was observed in overdoped $(Y_{0.7}Ca_{0.3})Ba_2Cu_3O_{7-\delta}$ (*Ca-YBCO*), and the pairing symmetry is consistent with $(d_{x^2-y^2}+s)$, with a significant *s*-component (>20%). Third, the presence of dilute spinless (*S=0*) impurities (such as $Zn^{2+}$ and $Mg^{2+}$) in *YBCO* resulted in microscopic spatial modulations in the quasiparticle spectra and strong suppression of superconductivity near the impurities. Furthermore, the global pairing potential is suppressed, suggesting long-range effects of the spinless impurities. Fourth, the doping dependence of the pairing potential $\Delta_d$ in *YBCO* is non-monotonic and is consistent with that of the superfluid density. Finally, the satellite features in the quasiparticle spectra of *YBCO* follow a doping dependence similar to that of $\Delta_d$. The physical significance of our studies and comparison of the *YBCO* system with *Bi-2212* are also discussed.

## 2. Spatial homogeneity and directionality of quasiparticle tunneling spectra

The technique used for this work involves a low-temperature scanning tunneling microscope (LT-STM) for studies of the directionality and spatial homogeneity of quasiparticle tunneling spectra in cuprate superconductors. The samples included: (1) three optimally doped *YBCO* single crystals, with $T_c$ = 92.9±0.1 K; (2) three underdoped *YBCO* single crystals with $T_c$ = 60.0±1.5 K; (3) One optimally doped and one underdoped $YBa_2Cu_3O_{7-\delta}$ c-axis epitaxial films, with $T_c$ = 91.0±1.0, and 85.0±1.0 K; (4) two overdoped *Ca-YBCO* epitaxial films with $T_c$ = 78.0±2.0 K; (5) one optimally doped $YBa_2(Cu_{0.9934}Zn_{0.0026}Mg_{0.004})_3O_{6.9}$ single crystal, hereafter denoted as (*Zn,Mg*)-*YBCO* with $T_c$ = 82.0±1.5 K. More details of our STM capabilities and the sample surface preparation procedure have been described elsewhere [16-18]. The spectra on single crystals were taken primarily with the quasiparticle tunneling direction $k$ parallel to three crystalline axes: the anti-node axes {100} or {010}, the nodal direction {110}, and the c-axis {001}. On the other hand, the spectra of *YBCO* and *Ca-YBCO* epitaxial films were taken with $k$ || {001}. Except the untwinned (*Zn,Mg*)-*YBCO* single crystal, all samples studied were twinned.

Figure 1 illustrates representative tunneling spectra at 4.2 K: (a) an optimally doped *YBCO* single crystal, with

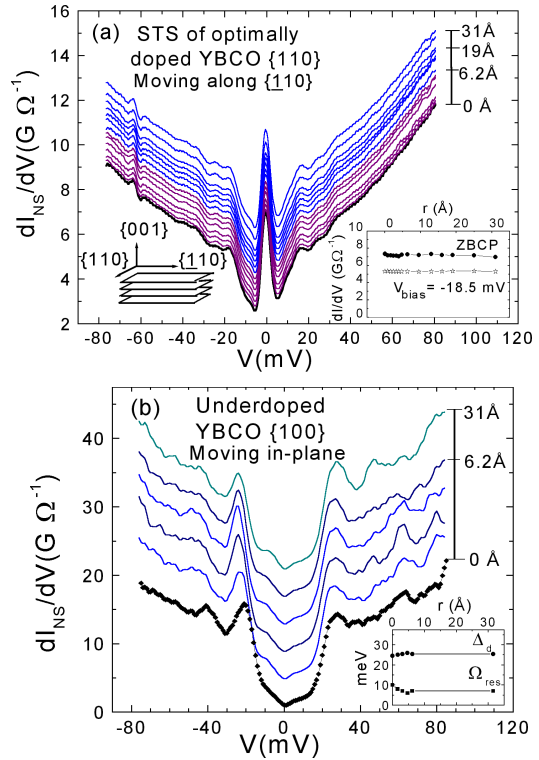

**Fig. 1** Representative tunneling conductance ($dI_{NS}/dV$) vs. bias voltage (*V*) data with atomic-scale spatial resolution at 4.2 K: (a) optimally *YBCO* ($T_c$ = 92.9 K), with $k$ || {110} and scanning along {$\bar{1}$10}; the inset demonstrates weak variations in the magnitude of the ZBCP and the satellite features; (b) underdoped *YBCO* ($T_c$ = 60.0 K), with $k$ || {100} and scanning along {010}; the inset shows weak spatial variation in the pairing potential ($\Delta_d$) and the satellite feature ($\Omega_{res}$). The spatial homogeneity extends up to ~ 100 nm, although the data here only focus on a shorter range for clarity.

quasiparticle momentum $k$ || {110} and the STM tip scanning along the {$\bar{1}$10} direction, as illustrated in the left inset; (b) an underdoped *YBCO* single crystal, with $k$ || {100} and scanning along {010} direction. (Additional spectra along other tunneling and scanning directions could be found in Ref. [18]). We find that these spectra exhibit long-range spatial homogeneity up to ~ *100 nm* for all tunneling and scanning directions, and are consistent with predominant $d_{x^2-y^2}$ (>95%) pairing symmetry, as analyzed using the generalized Blonder-Tinkham-Klapwijk (BTK) theory [16-20]. The spatial homogeneity of the zero-bias conductance peak (ZBCP) associated with the Andreev bound states in the nodal direction of a *d*-wave superconductor [19,20] is shown in the inset of Figure 1(a). Similarly, the right inset in Figure 1(b) demonstrates the spatial

homogeneity of the pairing potential $\Delta_d$ and the energy scale $\Omega_{res}$ for a satellite feature (see Section 4) that were derived from the tunneling spectra. The spatially homogeneous quasiparticle spectra therefore enabled unambiguous determination of the pairing symmetry and pairing potential. In contrast, microscopic spectra variations in *Bi-2212* [21] and macroscopic phase separations in overdoped $La_{2-x}Sr_xCuO_4$ (*LSCO*) [22] have been reported, suggesting complications in determining the pairing properties of those systems.

### 3. Doping-dependent pairing symmetry and pairing potential

In contrast to the quasiparticle spectra of underdoped and optimally doped *YBCO*, the c-axis tunneling spectra of overdoped *Ca-YBCO* revealed long-range symmetric "subgap" features that are consistent with the $(d_{x^2-y^2}+s)$ pairing symmetry, according to the generalized BTK analysis. In Figure 2(a) we compare the c-axis tunneling spectrum of an overdoped *Ca-YBCO* with that of an underdoped *YBCO*. The generalized BTK calculations for c-axis tunneling spectra of different pairing symmetries is shown in the inset of Figure 2(a), following the same procedure outlined in Refs. [16,17]. Defining $\xi_k$ as the wave-vector ($k$)-dependent single-particle energy relative to the Fermi level $E_F$, $\Delta_k$ and $E_k$ as the pairing potential and quasiparticle energy, with $E_k^2 = \Delta_k^2 + \xi_k^2$, and $Z$ being the tunneling barrier parameter, we obtain the tunneling current ($I_{NS}$) as a function of the bias voltage ($V$) [16-20]:

$$I_{NS} = G_{NN} \int exp[-(k_t/\beta)^2]\, d^2k_t$$
$$\times \int dE_k\, [1+A(E_k,\Delta_k,Z)-B(E_k,\Delta_k,Z)][f(E_k-eV)-f(E_k)].$$

Here $G_{NN}$ is proportional to the normal-state junction conductance, $A$ and $B$ are the Andreev-reflection and normal reflection probabilities, $f$ is the Fermi function for quasiparticle distributions, and $\beta$ is the tunneling cone for the effective spread of the transverse momentum ($k_t$) and the surface roughness [16-18]. To estimate the contributions from different pairing components, we consider $\Delta_k=\Delta_d\cos(2\theta_k)+i\Delta_s$ for $(d_{x^2-y^2}+is)$, $\Delta_k =\Delta_d\cos(2\theta_k)+i\Delta'\sin(2\theta_k)$ for $(d_{x^2-y^2}+id_{xy})$, and $\Delta_k=\Delta_d\cos(2\theta_k)+\Delta_s$ for $(d_{x^2-y^2}+s)$. As illustrated in the inset of Figure 2(a), the calculated c-axis tunneling spectra clearly distinguish the pairing symmetries that permit nodal quasiparticles (such as $d_{x^2-y^2}$ or $d_{x^2-y^2}+s$) from those with broken $\mathcal{T}$-symmetry (such as $d_{x^2-y^2}+is$ or $d_{x^2-y^2}+id_{xy}$). That is, the latter pairing state would have resulted in a fully gapped Fermi surface and the absence of density of states near $E_F$ at $T << T_c$. The distinct V-shaped spectra near $V = 0$ of the tunneling spectra for all *YBCO* samples indicated the existence of nodal quasiparticles for all doping levels. On the other hand, the symmetric subgap features over a long spatial range in the overdoped *Ca-YBCO* were consistent with the $(d_{x^2-y^2}+s)$ pairing, with a significant *s*-component.

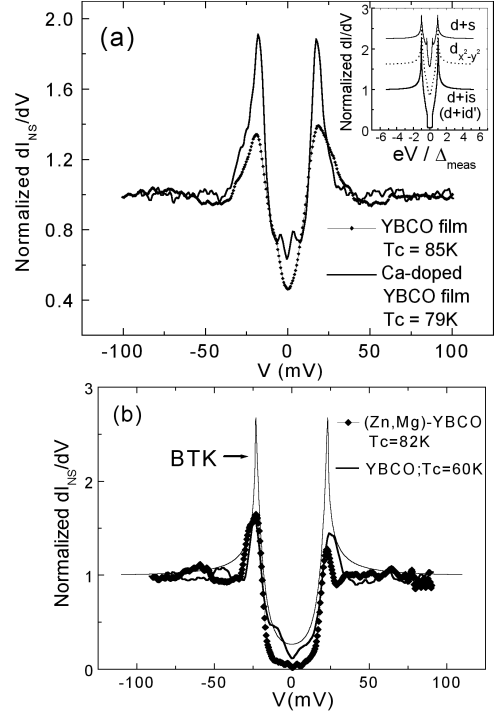

**Fig. 2** (a) Comparison of the c-axis tunneling spectrum of an overdoped *Ca-YBCO* ($T_c = 79$ K) with that of an underdoped *YBCO* ($T_c = 85$ K). The inset illustrates the generalized BTK calculations for c-axis tunneling spectra with $d_{x^2-y^2}$, $d_{x^2-y^2}+is$ ($d_{x^2-y^2}+id_{xy}$) and $(d_{x^2-y^2}+s)$ symmetries, using $\Delta_d = 19$meV, $\Delta_s = 6$meV. (b) The tunneling spectrum of *(Zn,Mg)-YBCO* ($T_c = 82$ K) for $k||\{100\}$ is compared with that of an underdoped *YBCO* ($T_c = 60$ K) and also with the BTK calculations.

For completeness, the $k \parallel \{100\}$ quasiparticle spectra of underdoped *YBCO* ($T_c = 60$ K) and *(Zn,Mg)-YBCO* are compared with the generalized BTK calculations in Figure 2(b). We note that the theoretical results predicted a U-shaped spectrum for $k$ along the anti-node direction. In real samples, however, residual quasiparticle states with microscopic spatial variation existed inside the U-shaped "gap". These states may be attributed to disorder-induced quasiparticle states, and the resulting V-shaped residual spectra suggested that nodal quasiparticles were the primary low-energy excitations induced by disorder. Under the premise of long-range spatial homogeneity in the pairing state, the doping-dependent pairing symmetry from $d_{x^2-y^2}$ to ($d_{x^2-y^2}+s$) with increasing doping level in *YBCO* is consistent with significant changes in the superconducting ground state beyond certain doping level, while the Fermi surface remains gapless along certain nodal directions.

In addition to the doping-dependent pairing symmetry, $\Delta_d$ in the *YBCO* system is non-monotonic with the doping level ($p$), as illustrated in Figure 3(a). This finding is in sharp contrast to the doping-dependence of the energy gap $\Delta^*$ observed in the *Bi-2212* system [23-25]. On the other hand, the ratio ($2\Delta_d/k_BT_c$) *increases* with decreasing $p$, and the values in underdoped samples significantly exceed that of the mean-field *d*-wave ratio ($2\Delta_d/k_BT_c$) ~ *4.3* [26]. The large ($2\Delta_d/k_BT_c$) ratio may be understood as follows. The magnitude of $\Delta_d$ represents zero-temperature pairing characteristics, whereas $T_c$ is associated with the phase coherence of the order parameter, and therefore can be strongly influenced by fluctuation effects. Thus, we expect more rapid decline in $T_c$ than in $\Delta_d$ with decreasing $p$.

## 4. Satellite features in the quasiparticle spectra

In addition to the primary peak features in the tunneling spectra that provide information for the pairing symmetry and pairing potential, we note that satellite features at higher energies are present in all *YBCO* samples. In Bi-2212, the "dip/hump" satellite features following a peak have been widely observed in various experiments such as the tunneling spectra [23,24] and angular-resolved photoemission spectroscopy (ARPES) [25,27]. Those features have been attributed to quasiparticle damping via interaction with collective spin fluctuations [28,29]. Under such a scenario, a "dip" in the spectral function would appear at the energy $\omega_0=\Delta_{meas}+\Omega_{res}$ in the strong coupling limit, where $\Omega_{res}$ is related to the resonance of propagating collective spin excitations, with $\Omega_{res} \sim (\Delta_{meas})^{1/2}v_F/(g\xi)$, $v_F$ being the Fermi velocity, $\xi$ the superconducting coherence length, and $g$ the coupling constant between quasiparticles and spin excitations [28]. Thus, if the measured gap $\Delta_{meas}$ is associated with the superfluid density, we expect $\Omega_{res}$ to decrease due to decreasing $\Delta_{meas}$ and increasing $g$ with decreasing doping level in the underdoped regime. On the other hand, both $\Delta_{meas}$ and $g$ are known to decrease with increasing $p$ in the overdoped limit, so that $\Omega_{res}(p)$ could only be determined empirically. By taking $\Omega_{res}$ as the energy difference between the primary peak and the dip in the spectra, we find that the doping dependence of $\Omega_{res}$ in *YBCO* system is qualitatively similar to that of $\Delta_d$, and is in contrast to the behavior of $\Delta^*(p)$ in *Bi-2212*, as shown in Figure 3(b). We note that additional peak-like satellite features at higher energies in the *YBCO* system (immediately following the spectral dip) differ from the broad "hump" feature in *Bi-2212*. Comparing the energy of the secondary peak ($\Omega_2$) in *YBCO* with $\Delta^*(p)$ in *Bi-2212*, we find that $\Omega_2$ of optimally doped *YBCO* is significantly larger than that of the underdoped *YBCO*, in contrast to the doping dependence of the pseudogap. Thus, the secondary peak may be the result of higher-order quasiparticle interactions with collective spin excitations rather than the pseudogap. Future spectral studies at $T > T_c$ will be necessary to resolve this issue.

## 5. Effects of spinless (S=0) impurities

In contrast to the long-range spatial homogeneity in the quasiparticle spectra of *YBCO* and *Ca-YBCO*, *microscopic* spatial variations have been observed in the (*Zn,Mg*)-*YBCO* single crystal near the *Zn* and *Mg* sites. It is worth noting that non-magnetic impurities (i.e. spinless with $S = 0$, such as $Zn^{2+}$ [30-33], $Mg^{2+}$ [34], $Li^+$ [35] and $Al^{3+}$ [36]) that substitute the $Cu^{2+}$-ions in the CuO$_2$ plane can induce local magnetic moments that significantly perturb the immediate vicinity of the impurity site, yielding suppression of superconductivity at $T < T_c$ and strong effects on the spin dynamics at $T > T_c$. On the other hand, magnetic impurities (such as $Ni^{2+}$ with $S = 1$) are coupled to the background spin fluctuations via exchange interaction, and therefore act as weaker and more local scattering sites [30,31]. The effects of spinless impurities on cuprates are fundamentally different from those on conventional superconductors [37], and have been attributed to strong-correlation phenomena [38]. Among various studies of the impurity effects on cuprates, the most revealing data have been the observation of strong spectral modulations near the site of non-magnetic impurities, such as the STM studies on *Zn*-doped *Bi-2212* [33], and our recent work on (*Zn,Mg*)-*YBCO* [18]. The essence of our finding is that at the impurity sites, the coherent quasiparticle peaks for the c-axis tunneling spectra are strongly suppressed and replaced by a single impurity scattering peak at an energy $|E_0|<<\Delta_d$. Furthermore, the quasiparticle spectra exhibit strong spatial modulations at the microscopic scale. The usual c-axis quasiparticle spectrum is recovered at several coherence lengths away from the nonmagnetic impurity site [18], and the global value of $\Delta_d$ is suppressed relative to that of the optimally doped YBCO, (see Figure 3(a)). The reduction in $\Delta_d$ is also consistent with the increase in the coherence length of Zn-doped YBCO [32]. The energy scale $\Omega_{res}$ associated with the usual spectral dip is also reduced significantly, as shown in Figure 3(b). These findings are consistent with the long-range effects of the spinless impurities, as inferred from other studies such as NMR and neutron scattering experiments [30,31].

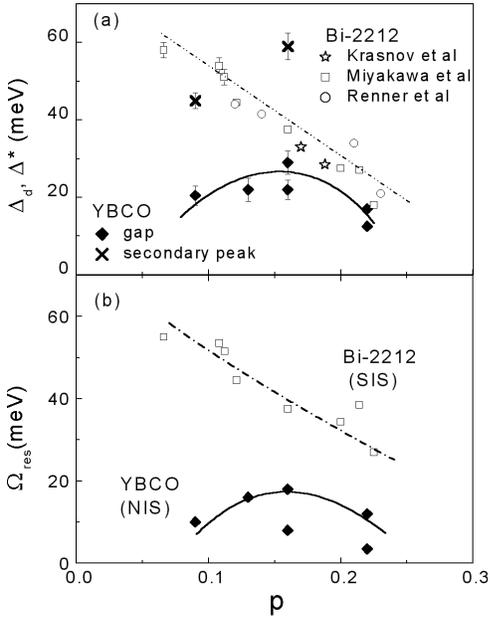

**Fig. 3** (a) $\Delta_d(p)$ of the YBCO system is compared with the average measured gap $\Delta^*(p)$ in Bi-2212. The doping level $p$, except for the optimally doped (Zn,Mg)-YBCO, is determined by the formula $T_c = T_{c,max}[1-82.6(p-0.16)^2]$, with $T_{c,max} = 93.0$ K for the optimally doped YBCO. The global value of $\Delta_d$ in the optimally doped (Zn,Mg)-YBCO is reduced relative to that of pure YBCO. (b) Comparison of $\Omega_{res}(p)$ and $\Omega_2(p)$ for YBCO and Bi-2212. Note the resemblance of $\Omega_{res}(p)$ to $\Delta_d(p)$.

In the context of pairing symmetry, we further note that the single scattering peak at a non-magnetic impurity site is not compatible with any broken $\mathcal{T}$-symmetry component $\Delta'$ in the pairing potential of YBCO, because the latter scenario would have resulted two excitation peaks at $\pm E_0$ [39]. In the strong impurity scattering limit, $|E_0|$ is given by $|E_0| \approx \Delta_d/[(\pi N_F V_{imp})\ln(4\Delta_d/\Delta')]$, and $|E_0|<<\Delta_d$ is satisfied because $\pi N_F V_{imp}>>1$, where $N_F$ is the density of states at the Fermi level, $V_{imp}$ the on-site impurity scattering potential, and $\Delta'<<\Delta_d$ has been assumed [39]. Hence, the absence of double peaks in the quasiparticle spectra for both (Zn,Mg)-YBCO [18] and Zn-doped Bi-2212 [33] systems provides additional confirmation for a gapless Fermi surface in the superconducting state of these cuprates.

## 6. Discussion

Our studies of quasiparticle tunneling spectra support the notion that nodal quasiparticles are the primary low-energy excitations of the cuprate superconductors, and that strong coupling to collective excitations exists in the underdoped and optimally doped samples. Concerning the possible existence of a quantum critical point (QCP) in the cuprates [3], our finding of doping-dependent pairing symmetry from predominantly $d_{x^2-y^2}$ to $(d_{x^2-y^2}+s)$ in the YBCO system suggests interesting changes in the superconducting ground state possibly at a critical doping level. However, we caution that the slight orthorhombicity of optimally doped YBCO may have given rise to a pairing symmetry that constitutes to a very small s-component beyond our experimental resolution. Thus, the doping dependent pairing symmetry may have been the result of a significant increase in the s-component in the overdoped limit, while preserving $\mathcal{T}$-symmetry. This conjecture is consistent with Raman scattering studies [40] where small in-plane anisotropy in optimally doped YBCO and rapidly vanishing anisotropy in underdoped YBCO has been reported. Noting that a naturally consequence of the $(d_{x^2-y^2}+s)$ pairing symmetry is the in-plane electronic anisotropy, the optical studies in Ref. [40] together with our tunneling experiments are suggestive of a pairing symmetry that evolves continuously from $d_{x^2-y^2}$ to $(d_{x^2-y^2}+s)$ with increasing doping. For comparison, the pairing symmetry in tetragonal Bi-2212 appears to be consistent with pure $d_{x^2-y^2}$ for all doping levels. Therefore our experiments do not directly support the existence of a QCP, although certain broken $\mathcal{T}$-symmetry states (such as the staggered flux state [41] or the circulating current phase [3]) or other hidden symmetries cannot be detected by quasiparticle tunneling spectra. Interestingly, our finding may be related to the presence of a *critical line*, rather than a critical point, in the underdoped regime of the $T$-$p$ phase diagram [13].

## 7. Summary

Studies of the low-temperature quasiparticle tunneling spectra of the YBCO system revealed long-range spatial homogeneity in the pairing potential $\Delta_d$ for a wide range of doping levels. This finding is in sharp contrast to the *microscopic* spatial variations in the quasiparticle spectra of *Bi-2212*, where the oxygen distribution is known to be random and differs from that in YBCO. The pairing symmetry was doping dependent, evolving from predominantly $d_{x^2-y^2}$ in the underdoped samples to ($d_{x^2-y^2}+s$) in the overdoped regime, suggesting significant changes in the ground state at certain doping level while always maintaining nodes on the Fermi surface. The doping dependence of $\Delta_d$ was non-monotonic, whereas the ($2\Delta_d/k_BT_c$) ratio increased with decreasing doping, suggesting stronger deviation from the mean-field behavior in the underdoped regime. The presence of spinless impurities resulted in strong suppression of superconductivity at a *microscopic* length scale near the impurities and global reduction in $\Delta_d$ over a *macroscopic* length scale. These results implied the importance of nodal quasiparticles and strong phase fluctuations in determining the physical properties of the cuprates.

**Acknowledgement**

The work at Caltech was supported by NSF, at Augsburg University by BMBF Grant #13N6918/1, and at SRL by the New Energy and Industrial Technology Development Organization (NEDO).